\documentclass[]{jd04}    
\usepackage[]{graphicx}

\Title[Two-shell SNRs]{On the origin of two-shell supernova
remnants} \Author{Gvaramadze, Vasilii}
\Address{Sternberg Astronomical Institute, Moscow State University,\\
Universitetskij Pr. 13, Moscow, 119992, Russia
(vgvaram@sai.msu.ru)\\}{}

\ShortAuthors{Gvaramadze}


\begin{document}
\maketitle

\begin{abstract}

The proper motion of massive stars could cause them to explode far
from the geometric centers of their wind-driven bubbles and thereby
could affect the symmetry of the resulting diffuse supernova
remnants (SNRs). We use this fact to explain the origin of SNRs
consisting of two partially overlapping shells (e.g. Cygnus Loop,
3C~400.2, etc.).
\end{abstract}

\section{Introduction}

The SNRs consisting of two partially overlapping nonthermal shells
represent a significant fraction of the known shell-like SNRs.
Several explanations were put forward to explain the unusual
morphology of these objects: {\it i}) superposition of two separate
SNRs; {\it ii}) collision of two separate SNRs (e.g. Uyaniker et al.
2002); {\it iii}) breakout phenomenon in a region with a density
discontinuity (e.g. Vel\'{a}zquez et al. 2001). We propose an
alternative explanation based on the idea that the two-shell SNRs
can be products of an off-centered supernova (SN) explosion in a
preexisting bubble created by the wind of a moving massive star.

\section{Precursors of two-shell SNRs}

Massive stars (the progenitors of most of SNe) are the sources of
strong stellar wind, which creates extended (tens of parsecs)
bubbles in the interstellar medium. In the absence of stellar proper
motion, the wind bubbles blown-up in the homogeneous medium are
spherically-symmetric with the star at the center of symmetry.
However, after correction for galactic rotation, all stars have a
proper motion. Most of the massive stars have a peculiar velocity of
a few km/s; some of them have much larger velocities. The stellar
proper motion could result in a considerable offset of the SN
explosion site from the center of the wind-driven bubble, while the
initially spherical shape of the bubble could be significantly
distorted if the star reaches the edge of the bubble and the stellar
wind starts to interact directly with the ambient interstellar
medium (Weaver et al. 1977; see Fig.~1 for schematic illustration of
this effect).

\begin{figure}[h]

\centering
\caption{Schematic illustration of the effect of stellar motion on
the structure of a wind bubble. At early times (A) the small circle
corresponds to a shock separating freely expanding stellar wind from
the region of shocked wind. The large circle corresponds to the
shock separating the unperturbed interstellar gas from the swept-up
(shocked) interstellar gas. (The contact discontinuity separating
the swept-up gas from the shocked wind is omitted for the sake of
simplicity.) At advanced times (B) the stellar wind interacts
directly with the ambient interstellar medium.} \label{fig1}
\end{figure}

It is likely that just this situation takes place in the case of the
progenitor star of the SN~1987A. Wang et al. (1993) suggested that
the large-scale structure to the southeast of SN~1987A (the dark bay
in their Fig.~2) is the wind-driven bubble created by the moving SN
progenitor star during the main-sequence (MS) stage, while the
Napoleon's Hat nebula originates due to the interaction of the
post-MS winds with the ambient interstellar medium (see, however,
Sugerman et al. 2005).

\begin{figure}[h]

\centering
\caption{The continuum-subtracted ${\rm H}_{\alpha}$ mosaic of
RCW~86 (Smith 1997).} \label{fig3}
\end{figure}

After the massive star exploded as a SN, the SN blast wave takes on
the shape of the preexisting cavity (e.g. R\'{o}$\dot{z}$yczka et
al. 1993; Brighenti \& D'Ercole 1994; Gvaramadze \& Vikhlinin 2003)
and the resulting SNR becomes considerably
non-spherically-symmetric. Fig.~2 shows the ${\rm H}_{\alpha}$ image
(Smith 1997) of RCW~86 -- a shell-like SNR with a peculiar
protrusion to the southwest. We believe (Gvaramadze \& Vikhlinin
2003) that RCW~86 is an older version of the SN~1987A and that the
southwest protrusion is the remainder of a bow shock-like structure
created in the interstellar medium by the post-MS winds. Given the
youth of the SNR, we expect that the stellar remnant should still be
within the protrusion. Motivated by these arguments we searched for
a stellar remnant using the {\it Chandra} archival data and
discovered a neutron star candidate just in the "proper place"
(Gvaramadze \& Vikhlinin 2003).

\section{Two-shell SNRs}

We propose that a two-shell SNR is a natural consequence of an
off-centered cavity SN explosion of a moving massive star, which
ended its evolution near the edge of the MS wind-driven bubble. This
proposal implies that one of the shells is the former MS bubble
reenergized by the SN blast wave. The second shell, however, could
originate in two somewhat different ways, depending on the initial
mass of the SN progenitor star. It could be a shell swept-up by the
SN blast wave expanding through the unperturbed ambient interstellar
medium if the massive star ends its evolution as a red supergiant
(RSG), i.e. if the star evolves through the sequence: MS-RSG-SN. Or
it could be the remainder of a preexisting shell (adjacent to the MS
bubble) swept-up by the fast progenitor's wind during the late
evolutionary phases if after the RSG phase the star evolves through
the Wolf-Rayet (WR) phase (i.e. MS-RSG-WR-SN). In both cases the
resulting (two-shell) SNR should be associated only with one (young)
NS (cf. Gvaramadze 2006). We note several distinctions
characterizing the second case: (a) the birth-place of the stellar
remnant could be significantly offset from the center of the
preexisting WR shell due to the proper motion of the SN progenitor
star, (b) the SN explosion site could be marked by a nebula of
thermal X-ray emission (see Sect.\,4.2), and (c) the preexisting WR
shell causes the rapid evolution of the SN blast wave from the
adiabatic phase to the radiative one.

\section{Two examples}

\subsection{Cygnus Loop (MS-RSG-SN)}

\begin{figure}[h]

\centering
\caption{Two shells of the Cygnus Loop: {\it ROSAT} image at 0.25
keV with overlaid polarization intensity contours (Uyaniker et al.
2002). A neutron star candidate (Miyata et al. 2001) is indicated by
a cross.} \label{fig5}
\end{figure}

The polarized intensity image of the Cygnus Loop by Uyaniker et al.
(2002) shows a prominent shell-like structure encompassing the
"break-out" region in the south of this SNR (see Fig.~3). The
geometric center of the shell nearly coincides with a neutron star
candidate discovered in X-rays by Miyata et al. (2001; indicated in
Fig.~3 by a cross). Uyaniker et al. (2002) believe that the Cygnus
Loop is actually two individual SNRs interacting with each other. An
alternative possibility is that the Cygnus Loop is the result of SN
explosion near the south edge of a cavity blown up by the SN
progenitor during the MS stage and that the SN precursor was a RSG
star (Gvaramadze 2003). This implies that the north (well-known)
shell of the Cygnus Loop corresponds to the former MS bubble
reenergized by the SN blast wave, while the newly-discovered (south)
shell is created by the interaction of the SN blast wave with the
unperturbed interstellar medium. Accordingly, we expect that only
one stellar remnant should be associated with both shells.

\begin{figure}[h]

\centering
\caption{{\it ASCA} image of 3C~400.2 (Yoshita et al. 2001).
Overlaid contours are the VLA image at 1.4 GHZ (Dubner et al.
1994).} \label{fig6}
\end{figure}

\subsection{3C~400.2 (MS-RSG-WR-SN)}

The SNR 3C~400.2 consists of two circular radio shells with the
centrally-filled thermal X-ray emission peaked on the region where
the radio shells overlap each other (Fig.~4; Dubner et al. 1994;
Yoshita et al. 2001) and therefore belongs to the class of
mixed-morpology SNRs (i.e. shell-like in the radio and
centrally-filled in the X-ray; Rho \& Petre 1998). The origin of
mixed-morphology SNRs is usually treated in the framework of White
\& Long's (1991) model of evaporation of embedded interstellar
cloudlets. An alternative explanation of the origin of the
centrally-filled thermal X-ray emission is that it is due to the
evaporation of dense circumstellar clumps (produced by the
interaction of the fast WR wind with the preceding slow RSG one;
Gvaramadze 2001, 2002). We therefore suggest that the SN precursor
was a WR star and that the northwest shell of 3C~400.2 is the former
WR shell. This suggestion can be supported by the fact that the
northwest shell has a bilateral appearance with the bilateral axis
parallel to the Galactic plane (see Gvaramadze 2004). If our
suggestion is correct, then one can expect that the stellar remnant
associated with 3C~400.2 should be within the northwest shell.


\end{document}